\documentclass[prl,twocolumn]{revtex4-1}
\usepackage{graphicx}
\usepackage{amsmath}
\usepackage{bm}

\begin{document}
\title{Free energy calculations of a proton transfer reaction by simulated tempering umbrella sampling first principles molecular dynamics simulations}
\author{Yoshiharu Mori}
\email[E-mail: ]{ymori@tb.phys.nagoya-u.ac.jp}
\affiliation{Department of Physics, Graduate School of Science, Nagoya University, Nagoya, Aichi 464-8602, Japan}
\author{Yuko Okamoto}
\email[E-mail: ]{okamoto@phys.nagoya-u.ac.jp}
\affiliation{Department of Physics, Graduate School of Science, Nagoya University, Nagoya, Aichi 464-8602, Japan}
\affiliation{Structural Biology Research Center, Graduate School of Science, Nagoya University, Nagoya, Aichi 464-8602, Japan}
\affiliation{Center for Computational Science, Graduate School of Engineering, Nagoya University, Nagoya, Aichi 464-8603, Japan}
\begin{abstract}
A new simulated tempering method, which is referred to as simulated tempering umbrella sampling, for calculating the free energy of chemical reactions is proposed.
First principles molecular dynamics simulations with this simulated tempering were performed in order to study the intramolecular proton transfer reaction of malonaldehyde in aqueous solution.
Conformational sampling in reaction coordinate space can be easily enhanced with this method, and the free energy along a reaction coordinate can be calculated accurately.
Moreover, the simulated tempering umbrella sampling provides trajectory data 
more efficiently than the conventional umbrella sampling method. 
\end{abstract}

\maketitle

Understanding chemical reactions by molecular simulations is a challenging problem, because a chemical reaction usually involves bond breaking, which cannot be treated by molecular simulations of classical mechanics based on force fields.
We need to use the first principles (or \textit{ab initio}) molecular dynamics (AIMD) methods to deal with bond breaking.
However, the time span that can be studied by the first principles molecular simulations is very much limited, because their computational cost is much higher than that of the simulations with force fields.

The molecular simulations based on quantum or classical mechanics are hampered by the multiple-minimum problem, in which simulations tend to get trapped in the 
local-minimum states of free energy.
One can use generalized-ensemble algorithms to overcome this difficulty (for a reviews, see, e.g., \cite{Mitsutake01}).
Monte Carlo (MC) and molecular dynamics (MD) simulations based on generalized-ensemble algorithms have been widely performed for many molecular systems in order to have efficient conformational sampling.
Three well-known generalized-ensemble algorithms are the multicanonical algorithm (MUCA)~\cite{Berg91,Berg92} (for the MD version see 
Refs.~\cite{Hansmann96,Nakajima97}), the replica-exchange 
method (REM)~\cite{Hukushima96} (the method is also referred to as 
parallel tempering~\cite{Marinari98} and for the MD version, 
which is referred to as REMD, see Ref.~\cite{Sugita99}), and 
simulated tempering (ST)~\cite{Lyubartsev92,Marinari92}. 

Recently, general formulations for the multidimensional MUCA, REM, and ST 
have been given~\cite{Mitsutake09a,Mitsutake09b,Mitsutake09c}.
In this article we introduce a special realization of the generalized ST, 
which we refer to as Simulated Tempering Umbrella Sampling (STUS).
This is a generalization of the Umbrella Sampling (US) method~\cite{Torrie77} 
and is closely related to the Replica-Exchange Umbrella Sampling (REUS) 
in Ref. \cite{Sugita00a}.

Let us consider a system that consists of $N$ atoms, where atom $i$ has 
coordinate $\bm{r}_i$.
We write the set of the coordinates of the atoms 
as $r = \{\bm{r}_1, \bm{r}_2, \cdots, \bm{r}_N\}$.
The original potential energy function is represented by $E(r)$, 
and the temperature by $T$.

We propose a ST method in parameter space.
The parameter here stands for a label that specifies the potential energy 
function.
In this STUS method, $M$ different restraint potential energy
functions (or, umbrella potentials) $V_1, V_2, \cdots, V_M$ are used.
Introducing a reaction coordinate $\xi$, we define the umbrella
potential function $V_m$ by
\begin{equation}
\label{eq:umbrella}
V_m\left( \xi(r)\right) = k_m\left[\xi(r) - d_m  \right]^2~, 
\quad (m=1,2,\cdots,M),
\end{equation}
where $k_m$ are force constants and $d_m$ are equilibrium distances of 
the reaction coordinate.
The STUS simulation yields a uniform probability distribution in parameter 
(or label) space.
It means that a random walk in the $M$ different umbrella
potential functions is realized during the simulation.

In the STUS method, each state is specified by coordinate $r$ and 
label $m$, and the following probability distribution function $W_m$ is used:  
\begin{eqnarray}
W_m(r) &=& \mathcal{Z}^{-1}\exp \left\{ -\beta 
\left[ E(r) + V_m(r)\right] + a_m \right\}, \nonumber \\  
&~&~~~~~~~~~~~~\quad (m=1,2,\cdots,M),
\end{eqnarray}
where $\beta$ ($=1/k_{\text{B}} T$) is the inverse temperature 
($k_{\text{B}}$ is the Boltzmann constant),
$\mathcal{Z}$ is defined by
\begin{equation}
\mathcal{Z} = \sum_{m=1}^{M}\int dr \ \exp 
\left\{ -\beta \left[ E(r) + V_m(r) \right] + a_m\right\},
\end{equation}
and $a_m$ are introduced so that the probability distribution 
in parameter space may be uniform.
If the probability in parameter space is constant, $a_m$ 
are formally written as the following dimensionless free energy 
except for a constant:
\begin{equation}
a_m = -\ln \left\{ \int dr \ \exp \left\{ -\beta \left[ E(r) + V_m(r) \right] \right\} \right\}.
\end{equation}

We update the umbrella potential function to 
another one every few steps during the STUS simulation.
When we attempt to change the $m$-th umbrella potential function $V_m$ to 
the $n$-th umbrella potential function $V_n$, we can use the following 
transition probability $w$ so that the detailed balance condition may be 
satisfied:
\begin{equation}
\label{eq:trans1}
w(m \to n)  = \min \left( 1, \frac{W_n}{W_m} \right) = \min \left( 1, \exp(-\varDelta) \right),
\end{equation}
where
\begin{equation}
\label{eq:trans2}
\varDelta = \beta \left[ V_n(r) - V_m(r)\right] - (a_n - a_m).
\end{equation}
We have used the Metropolis criterion~\cite{Metropolis53} to satisfy the detailed balance condition.

The ST parameters $a_m$ ($m=1, 2, \cdots, M$) can be determined by the 
(multiple-histogram) reweighting techniques applied to a preliminary 
replica-exchange simulation \cite{Mitsutake09a,Mitsutake09b,Mitsutake09c}.
Once the ST parameters are determined, the STUS simulations are performed by repeating the following two steps:
(1) perform a usual molecular simulation with the potential energy function
$E + V_m$ for 
some steps, (2) update the umbrella potential $V_m$ to a ``neighboring'' 
umbrella potential $V_n$ by the transition probability in 
Eqs. (\ref{eq:trans1}) and (\ref{eq:trans2}).
If accepted, replace $V_m$ by $V_n$.  Go back to step (1).

The free energy as a function of the reaction coordinate $\xi$, 
or the potential of mean force (PMF), $\mathcal{F}(\xi_0)$ for the original,
unbiased system is defined by
\begin{equation}
\label{eq:pmf1}
\mathcal{F}(\xi_0) = -k_{\text{B}}T\ln \left\{ Z^{-1}\int dr \ \delta\left( \xi(r)-\xi_0\right) \exp[-\beta E(r)]\right\},
\end{equation}
where $Z$ is the partition function at temperature $T$, and $\delta$ is the delta function.
Note that the umbrella potential functions are not included in this equation.

We can use another expression for PMF instead of Eq. (\ref{eq:pmf1}):
\begin{equation}
\mathcal{F}(\xi) = -k_{\text{B}}T\ln P(\xi) - C,
\end{equation}
where $P(\xi)$ is the probability distribution of the reaction 
coordinate $\xi$ in the original system without umbrella potential functions, 
and $C$ is an arbitrary constant to set the zero point of the free energy.
This form is convenient for molecular simulations because the 
probability $P(\xi)$ can be obtained by generating a histogram of 
the reaction coordinate.

From the results of the STUS simulation with the umbrella potential
functions, we can calculate $P(\xi)$ using some reweighting techniques 
such as the multistate Bennett acceptance ratio (MBAR) estimator~\cite{Shirts08}, which is based on the equations 
in Refs. \cite{Ferrenberg89} and \cite{Kumar92}.

The MBAR equations for calculating the expectation value $\langle A \rangle$ of a physical quantity $A$ are written as follows:
\begin{widetext}
\begin{eqnarray}
\label{eq:mbar1}
\langle A \rangle &=& \sum_{n=1}^{M}\sum_{k=1}^{N_n}
\frac{A(r_{n}(k))\exp [f - \beta E(r_{n}(k))]}
{\displaystyle \sum_{m=1}^{M}N_{m}\exp \{f_{m}-\beta [E(r_{n}(k))+V_m(r_{n}(k))]\}}, 
\\
\label{eq:mbar2}
f &=& - \ln \sum_{n=1}^{M}\sum_{k=1}^{N_n}
\frac{\exp [ -\beta E(r_{n}(k)) ]}
{\displaystyle \sum_{m=1}^{M}N_{m}\exp \{f_{m}-\beta [E(r_{n}(k))+V_m(r_{n}(k))]\}}, 
\\
\label{eq:mbar3}
f_{l} &=& - \ln \sum_{n=1}^{M}\sum_{k=1}^{N_n}
\frac{\exp \{ -\beta [E(r_{n}(k))+V_{l}(r_{n}(k))]\} }
{\displaystyle \sum_{m=1}^{M}N_{m}\exp \{f_{m}-\beta [E(r_{n}(k))+V_m(r_{n}(k))]\}},  
~~(l=1,2,\cdots,M),
\end{eqnarray}
\end{widetext}
where $N_m \ (m=1,2,\cdots,M)$ are the total numbers of trajectory data 
in the simulation with $V_m$, and $r_{n}(k)$ are the $k$-th coordinate
data in the trajectory obtained with $V_n$.
We can obtain the probability distribution $P(\xi$) of $\xi$ by calculating 
the expectation values of the number of $\xi$ taking the value 
$\xi_i \ (i=1,2,\cdots)$, where $\xi_i$ are discretized values 
along the reaction coordinate $\xi$.

\begin{figure*}
\begin{center}
\includegraphics[width=10cm,clip]{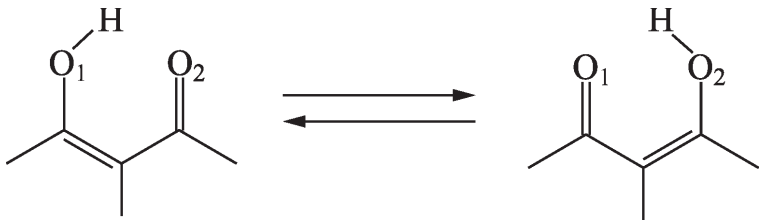}
\end{center}
\caption{Intramolecular proton transfer reaction of malonaldehyde. The hydrogen atom that transfers between the two oxygen atoms is written as $\text{H}$. The two oxygen atoms that can bond the hydrogen atom are written as $\text{O}_1$ and $\text{O}_2$.}
\label{fig:malon}
\end{figure*}

As an application of the STUS method, we considered the intramolecular proton transfer reaction of malonaldehyde (see Fig. \ref{fig:malon}).
We performed STUS AIMD simulations of malonaldehyde in order to show that 
the STUS method is effective in calculating the free energy of chemical reactions.

We defined the reaction coordinate $\xi$ of the proton transfer as the 
difference of two distances between the hydrogen atom $\text{H}$ and the two 
oxygen atoms $\text{O}_1$ and $\text{O}_2$:
\begin{equation}
\xi(r) = |\bm{r}_{\text{O}_1} - \bm{r}_{\text{H}}| - |\bm{r}_{\text{O}_2} - \bm{r}_{\text{H}}|.
\end{equation}
When malonaldehyde is in a stable state where hydrogen atom H bonds to oxygen atom $\text{O}_1$, $|\bm{r}_{\text{O}_1} - \bm{r}_{\text{H}}|$ should be less than $|\bm{r}_{\text{O}_2} - \bm{r}_{\text{H}}|$ and therefore $\xi < 0$.
Likewise, when hydrogen atom H bonds to oxygen atom $\text{O}_2$, $\xi > 0$.
When the reaction coordinate has a value which is nearly equal to 0, malonaldehyde should be in a transition state of the proton transfer reaction.

We used 11 different umbrella potential functions in Eq. (\ref{eq:umbrella}), that is, $M=11$.
$k_m$ were all set to be 0.01 hartree$\cdot$bohr$^{-2}$, 
and $d_1,d_2,\cdots,d_{11}$ were equally spaced between $-1.0$ \AA~ 
and $1.0$ \AA, namely, they were set 
to be $-1.0$ \AA, $-0.8$ \AA, $\cdots$, 1.0 \AA, respectively. 

We prepared the system of malonaldehyde in 71 water molecules with 
periodic boundary conditions. 
We used the CP2K program (version 2.1)~\cite{CP2K} to perform AIMD simulations 
based on the density functional theory with the Born-Oppenheimer approximation.
In the density functional calculations, we used the Becke exchange 
functional~\cite{Becke88} and the Lee-Yang-Parr correlation 
functional~\cite{Lee88}.
The pseudo potential proposed by Goedecker, Teter, and 
Hutter~\cite{Goedecker96,Hartwigsen98} was used.
A 280 Ry density grid was employed.
We used the mixed Gaussian and plane waves approach~\cite{Lippert97}.
We carried out the molecular simulations in the canonical ensemble.
The simulation cell was set to be a cubic box 
(13.82 \AA $\times$13.82 \AA $\times$13.82 \AA).
The temperature of the simulation system was set to be 300 K.
We used the Nos\'{e}-Hoover chain method~\cite{Nose84a,Hoover85,Martyna92} as a constant temperature algorithm, set the number of chains to be 3, and set the 
time constant of the Nos\'{e}-Hoover chain method to be 10 fs.
The time step of these simulations was set to be 0.5 fs.
The umbrella potential function was changed during the STUS simulation 
following the transition probability in Eqs. (\ref{eq:trans1}) 
and (\ref{eq:trans2}).
In the STUS simulation we attempted to update the umbrella potential function 
to a neighboring one every 10 steps, that is, every 5 fs.
We performed four independent simulations with the same conditions 
described above except for the initial velocities of the atoms.
The data were stored each time just after the ST update attempts. 
After discarding thermalization steps, we obtained 214 ps simulation data
altogether from the four STUS simulations.

In order to show that the STUS method is more efficient than the 
conventional US method, we also carried out AIMD US simulations with a 
fixed umbrella potential function using the same 11 umbrella potential 
functions as in the STUS simulations.
Each of the 11 US simulations was performed for 20 ps and the data were 
stored every 5 fs.
Actually, the ST parameters $a_m$ for the above STUS simulations
were determined by applying the MBAR
reweighting techniques to the results of these 11 US simulations.

\begin{figure*}
\begin{center}
\includegraphics[width=14cm,clip]{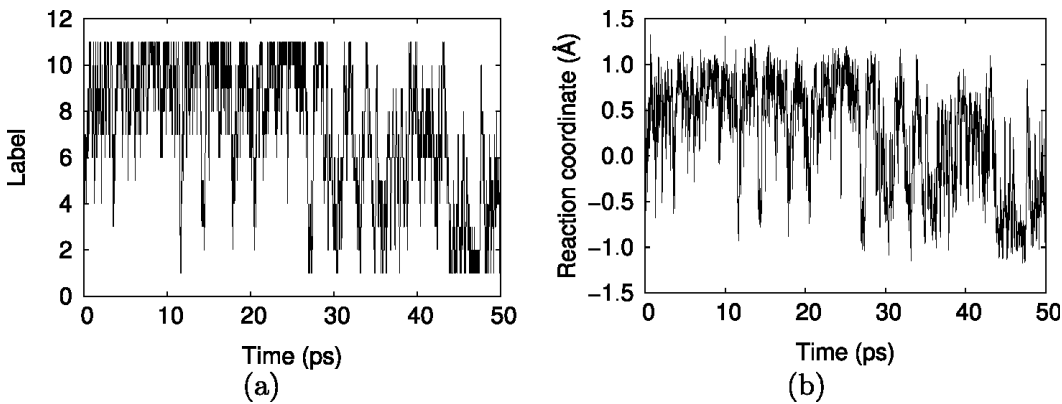}
\end{center}
\caption{Time series of (a) the label of the umbrella potential functions and (b) the reaction coordinate $\xi$ during the STUS molecular dynamics simulations.}
\label{fig:react}
\end{figure*}
The STUS method realizes a random walk in parameter space during the simulation.
As a result, the reaction coordinate corresponding to the parameter can be 
sampled much more widely
than in a conventional molecular simulation.
Figure \ref{fig:react} shows the time series of the label of the umbrella potential functions and those of the reaction coordinate $\xi$ in one of the four STUS AIMD simulations.
Through the simulation, the label of the umbrella functions largely 
fluctuated between 1 and 11 with almost equal probability.
The reaction coordinate also fluctuated 
greatly,
following the change of the umbrella potential.
Note that there is an expected strong correlation between the two graphs.
The reaction coordinate took on values between around $-1.5$ \AA ~and 1.5 \AA.
This means that malonaldehyde was able to experience not only the stable states, 
where hydrogen atom H bonds to either oxygen atom, but also the transition state.
Essentially the same results were also obtained for the other three simulations.

\begin{figure*}
\begin{center}
\includegraphics[width=14cm,clip]{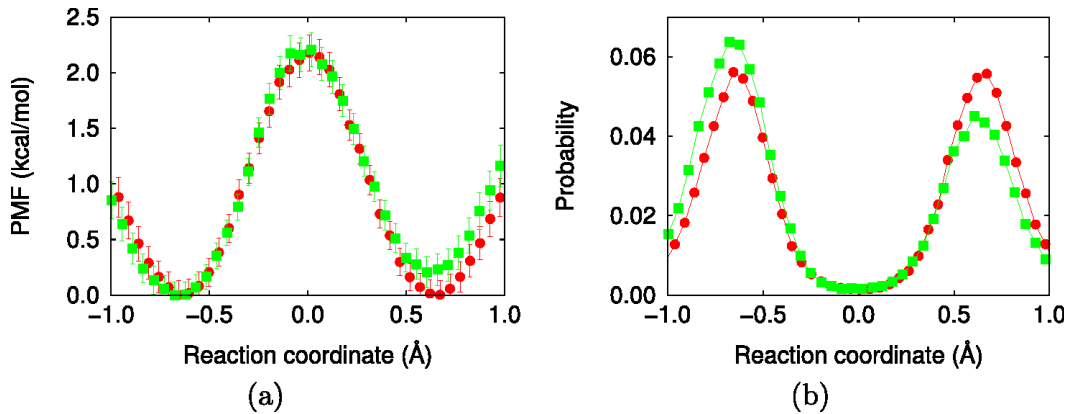}
\end{center}
\caption{(color online). (a) Potential of mean force (PMF) of the proton 
transfer reaction of malonaldehyde. The red filled circles represent
the PMF obtained by 
the STUS simulations and the green filled squares
by the conventional US simulations.  
The error bars were calculated by the MBAR 
estimator~\cite{Shirts08}. (b) Probability distribution of the reaction 
coordinate $\xi$. The red curve is the probability obtained by the 
STUS simulations and the green curve is that by the conventional
US simulations. 
The error bars are suppressed to aid the eye.}
\label{fig:pmf}
\end{figure*}
From the trajectory data of the STUS AIMD simulations, we calculated 
the PMF and the probability distribution of the reaction coordinate 
using the MBAR estimator in Eqs. (\ref{eq:mbar1}), (\ref{eq:mbar2}), 
and (\ref{eq:mbar3}), which are shown in Fig. \ref{fig:pmf}.
The PMF and probability distribution from the results of the 
conventional AIMD US simulations are also shown in Fig \ref{fig:pmf}. 

One would expect that the free energy of the proton transfer reaction of malonaldehyde is a double-well function which has minima at the two stable states and a local maximum at the transition state.
Figure \ref{fig:pmf} (a) shows that the PMF obtained from the STUS simulations has a simple double-well function which has two minima at nearly $\xi = -0.6$ \AA ~and $\xi = 0.6$ \AA, and a local maximum at around $\xi = 0.0$ \AA.

While the results of the STUS AIMD simulations were able to provide 
accurate PMF, the PMF calculated from the 
conventional US AIMD simulations could not be obtained accurately.
We can understand this clearly comparing the probability distribution
of $\xi$ obtained by the STUS simulations to that by the US simulations.
Although the probability in the two stable states is nearly equal to each other 
in the STUS simulations, the probability obtained by the US simulations  
is not equally distributed in the two stable states (see Fig. \ref{fig:pmf} (b)).
This implies that the US AIMD simulations could not sample sufficient 
trajectory data to calculate PMF in the time scale of the present 
simulations (20 ps for each simulation).

In summary, we have proposed a new simulated tempering method, Simulated Tempering Umbrella Sampling (STUS), and applied it to the first principles molecular dynamics simulations of the intramolecular proton transfer reaction of malonaldehyde.
We were able to obtain an accurate potential of mean force of the proton transfer reaction of malonaldehyde from the results of the STUS simulations.
We also compared the potential of mean force obtained by the STUS simulations with the one obtained by the conventional US simulations.
The STUS method is more efficient in exploring reaction coordinate space than the usual US method. 

In the present version of STUS, we fixed the temperature during the simulation.
We can easily generalize STUS so that a two-dimensional random walk in both temperature and umbrella potential, as was done in REUS \cite{Sugita00a}.

Moreover, the STUS method can be easily implemented in the existing program 
packages.
One does not need to modify the existing program.
Because only the difference of energy is needed in the simulation, one has 
only to write a simple script that extracts values of energy and 
reaction coordinates and evaluates the ST transition probability
during the simulation.

Some of the computations were performed on the supercomputers at the Information Technology Center, Nagoya University, at the Research Center for Computational Science, Institute for Molecular Science, and at the Supercomputer Center, Institute for Solid State Physics, University of Tokyo.
This work was supported, in part, by Grants-in-Aid for JSPS Fellows, for Scientific Research on Innovative Areas (``Fluctuations and Biological Functions"), and for the Computational Materials Science Initiative from the Ministry of Education, Culture, Sports, Science and Technology (MEXT), Japan.

\bibliography{myref}

\end{document}